\documentclass[twocolumn,showpacs,amsmath,amssymb,superscriptaddress]{revtex4-1}

\usepackage{graphicx}
\usepackage{dcolumn}
\usepackage{bm}
\usepackage[colorlinks=true,hyperindex=true,linkcolor=black,citecolor=black,urlcolor=black]{hyperref}

\begin{document}

\title{Interplay of Polarization and Time-Reversal Symmetry Breaking \\in Synchronously Pumped Ring Resonators}

\author{Fran\c cois Copie}
\email{francois.copie@npl.co.uk}
\affiliation{National Physical Laboratory (NPL), Teddington TW11 0LW, UK}
\author{Michael T. M. Woodley}
\affiliation{National Physical Laboratory (NPL), Teddington TW11 0LW, UK}
\affiliation{Heriot-Watt University, Edinburgh, EH14 4AS Scotland, UK}
\author{Leonardo Del Bino}
\affiliation{National Physical Laboratory (NPL), Teddington TW11 0LW, UK}
\affiliation{Heriot-Watt University, Edinburgh, EH14 4AS Scotland, UK}
\author{Jonathan M. Silver}
\affiliation{National Physical Laboratory (NPL), Teddington TW11 0LW, UK}
\author{Shuangyou Zhang}
\affiliation{National Physical Laboratory (NPL), Teddington TW11 0LW, UK}
\author{Pascal Del'Haye}
\email{pascal.delhaye@npl.co.uk}
\affiliation{National Physical Laboratory (NPL), Teddington TW11 0LW, UK}

\begin{abstract}
Optically induced breaking of symmetries plays an important role in nonlinear photonics, with applications ranging from optical switching in integrated photonic circuits to soliton generation in ring lasers. In this work we study for the first time the interplay of two types of spontaneous symmetry breaking that can occur simultaneously in optical ring resonators. Specifically we investigate a ring resonator (e.g. a fiber loop resonator or whispering gallery microresonator) that is synchronously pumped with short pulses of light. In this system we numerically study the interplay and transition between regimes of temporal symmetry breaking (in which pulses in the resonator either run ahead or behind the seed pulses) and polarization symmetry breaking (in which the resonator spontaneously generates elliptically polarized light out of linearly polarized seed pulses). We find ranges of pump parameters for which each symmetry breaking can be independently observed, but also a regime in which a dynamical interplay takes place. Besides the fundamentally interesting physics of the interplay of different types of symmetry breaking, our work contributes to a better understanding of the nonlinear dynamics of optical ring cavities which are of interest for future applications including all-optical logic gates, synchronously pumped optical frequency comb generation, and resonator-based sensor technologies.
\end{abstract}

\maketitle

Passive nonlinear optical cavities have been studied extensively in the past decades, partly for their ability to increase the efficiency of light-matter interactions through a large enhancement of circulating power \cite{grelu_nonlinear_2016}. Quite recently, the interest was renewed after the first observation of so-called cavity solitons (stable pulses of light circulating inside a resonator indefinitely) in macro-scaled fiber loops \cite{leo_temporal_2010} and microresonators \cite{herr_temporal_2014}, underpinning the generation of Kerr frequency combs \cite{delhaye_optical_2007}. In a number of practical studies, such systems are not driven by a continous wave (cw) laser but rather pumped by a train of pulses so that comparatively greater input peak powers are achieved \cite{copie_competing_2016, bendahmane_coherent_2017, wang_stimulated_2018} or to generate solitons and frequency combs with an improved efficiency \cite{obrzud_temporal_2017}. This however requires a rigorous control of either the pulse train repetition rate or cavity length to ensure the synchronicity of the pumping, the lack of which might alter the dynamics of the system \cite{coen_convection_1999, parra-rivas_effects_2014}.

Several studies have focussed on a scenario where the input pulses are Gaussian and their duration is longer than that of a typical cavity soliton. In that case, it has been observed that, provided that the resonator exhibits anomalous dispersion, the peak of the intracavity pulse does not necessarily lock at an extremum of the input power (symmetric solution). Instead, a solution where the peak of the soliton is shifted with respect to the extremum seems to be favoured. This phenomenon is referred to as a spontaneous symmetry breaking of the temporal pulse profile \cite{garcia-mateos_optical_1995, xu_experimental_2014, schmidberger_multistability_2014, rossi_spontaneous_2016} and has been very recently identified in the context of cavity soliton dynamics as resulting from a competition between the synchronous coherent driving and the nonlinear propagation inside the cavity \cite{hendry_spontaneous_2018}.

On a different note, although a large fraction of the work done on nonlinear resonators addresses the case of one-dimensional and single polarization propagation, polarization-related effects can greatly widen the range of phenomena occuring in such systems related, for instance, to instabilities \cite{haelterman_pure_1994}, pattern formation \cite{geddes_polarisation_1994, westhoff_pattern_2000, ackemann_polarization_2001}, soliton \cite{averlant_coexistence_2017} and frequency comb generation \cite{hansson_modulational_2018} or symmetry breaking between the different polarization modes \cite{areshev_polarization_1983, haelterman_polarization_1994, garcia-mateos_passive_1997, westin_polarization_1998, delque_symmetry_2008}. The latter item can be exploited for all-optical data transmission and storage, consecutive bits being connected by robust polarization domain walls \cite{fatome_polarization_2016, garbin_buffering_2018}. This can be achieved in the regime of normal dispersion where the formation of domain walls does not compete with the scalar process of modulation instability (MI). 

In the present work, we consider a system that supports both time-reversal (or temporal) and polarization symmetry breaking mechanisms: an isotropic ring cavity synchronously pumped by short pulses in the anomalous dispersion regime. We show by means of numerical simulations of a system of two coupled equations that when the detuning is scanned through the resonance of the cavity both symmetries can spontaneously be broken. We study the influence of the pump parameters (peak power and pulse duration) and focus on a configuration which enables a dynamical interplay between the two processes. The impact of power noise conditions on this interplay is also addressed. This work brings further insight into the actively studied complex dynamical behaviour of nonlinear optical resonators.

\begin{figure}[htbp]
	\centering
	\includegraphics[width=\linewidth]{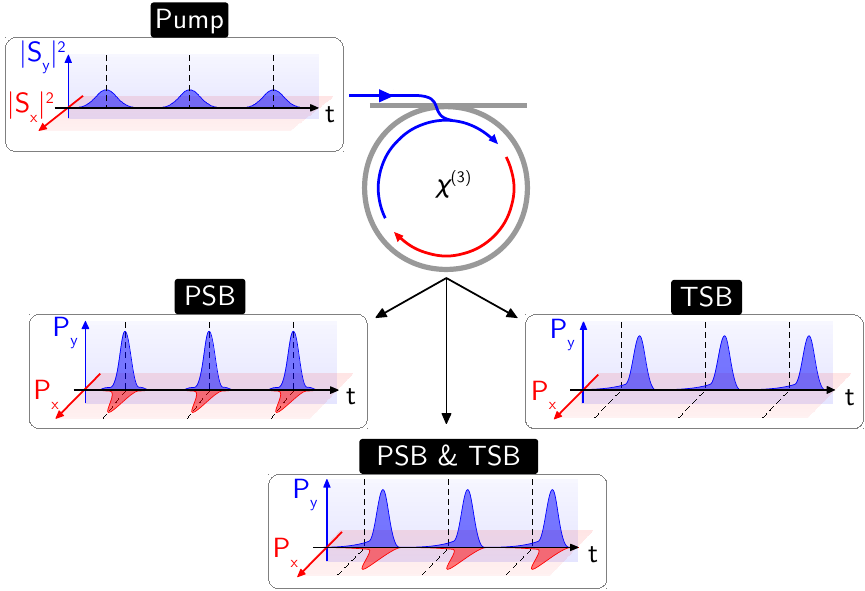}
	\caption{Schematic of the different types of symmetry breaking in a dielectric ring resonator. By convention, the pump field is linearly polarized along the $y$-axis. PSB: polarization symmetry breaking, TSB: temporal symmetry breaking.}
	\label{fig:1}
\end{figure}

We consider a passive ring cavity made of a dispersive medium exhibiting a Kerr nonlinearity, schematically represented in Fig.~\ref{fig:1}. Evolution of the intracavity field envelope in such a system is known to be well described by a one-dimensional Lugiato-Lefever equation (LLE) provided that (i) detuning from the resonance and round-trip losses are small (high finesse), (ii) fields evolve over a single transverse mode, and (iii) no polarization-related effects occur. In this work, we investigate the coupling between two counter-rotating circularly polarized modes inside the resonator. In that case, the evolution of the two fields over consecutive round-trips can be described by the following set of two normalized coupled Lugiato-Lefever equations (LLEs) \cite{haelterman_polarization_1994}:

\begin{equation}
\begin{split}
\frac{\partial E_\pm}{\partial z} - i \left[ \frac{1 - B}{2} |E_\pm|^2 + \frac{1 + B}{2} |E_\mp|^2 - \frac{\eta}{2}\frac{\partial^2}{\partial \tau^2}\right] E_\pm \\ + (1 + i\Delta) E_\pm = S_\pm(\tau)
\end{split}
\label{eq:LLEc}
\end{equation}

\noindent where $E_{\pm}$ ($S_{\pm}$) is the left/right circularly polarized component of the intracavity (input) field envelope respectively, $z$ is the unfolded longitudinal coordinate along the ring, $\tau$ is the fast time defined in the reference frame traveling at the group velocity of the pump, $\eta$ refers to the sign of the group velocity dispersion (+1 for normal dispersion; $-1$ for anomalous dispersion), and $\Delta$ is the cavity detuning (In this notation, $\Delta$ is expressed in units of half the resonance's linewidth at half maximum). The constant $B$ is related to the isotropic nonlinear medium of the cavity. It characterizes the ``strength'' of the coupling between the two fields of different polarization and we set it here to a value of $1/3$ which leads to the cross-phase modulation terms being twice as strong as the self-phase modulation terms \cite{haelterman_polarization_1994}. Note that any positive value would lead to qualitatively similar observations to the ones described in this work. Additionally, we assume here that both fields experience equal losses, detuning, and pump power. In the rest of this paper, we focus exclusively on the case of anomalous dispersion ($\eta = -1$) as it is a condition for the occurrence of temporal symmetry breaking \cite{garcia-mateos_optical_1995, xu_experimental_2014}. See Supplemental Material for details regarding the derivation of Eq.~(\ref{eq:LLEc}). The link between the two circularly polarized components and the linearly polarized ones is given by the following combinations:

\begin{equation}
	E_\pm = \frac{E_y \pm i E_x}{\sqrt{2}}~~~;~~~S_\pm = \frac{S_y \pm i S_x}{\sqrt{2}}
	\label{eq:orth}
\end{equation}

\noindent such that under a polarization-symmetric driving (i.e. $S_+ = S_-$) a symmetric state of the system (i.e. $E_+ = E_-$) corresponds to an intracavity field collinearly polarized with the pump (along the $y$-axis according to our notation). The polarization symmetry breaking (power imbalance between the circularly polarized components) thus manifests itself by the generation of an intracavity field component orthogonally polarized with respect to the pump. This is illustrated by the panel labelled ``\textit{PSB}'' (polarization symmetry breaking) in Fig.~\ref{fig:1}. In all the results presented here we consider Gaussian shaped pump field envelopes $S_\pm(\tau) = S_0 \exp[-(\tau/\tau_0)^2]$ identical for both circular polarizations except for the random noise that we include on each field. We investigate a range of pump parameters ($\tau_0,\, S_0$) limited by the condition that the intracavity field should not break into multiple pulses as a result of the MI process. This amounts to limiting the pump pulse duration to values shorter than the typical MI period. Moreover, this period depends itself on the intracavity power (the larger the power, the shorter the MI period) such that we also limit the investigation to rather low pump peak power values (i.e. close to the symmetry breaking thresholds as introduced later).

\begin{figure*}[htbp]
	\centering
	\includegraphics[width=\linewidth]{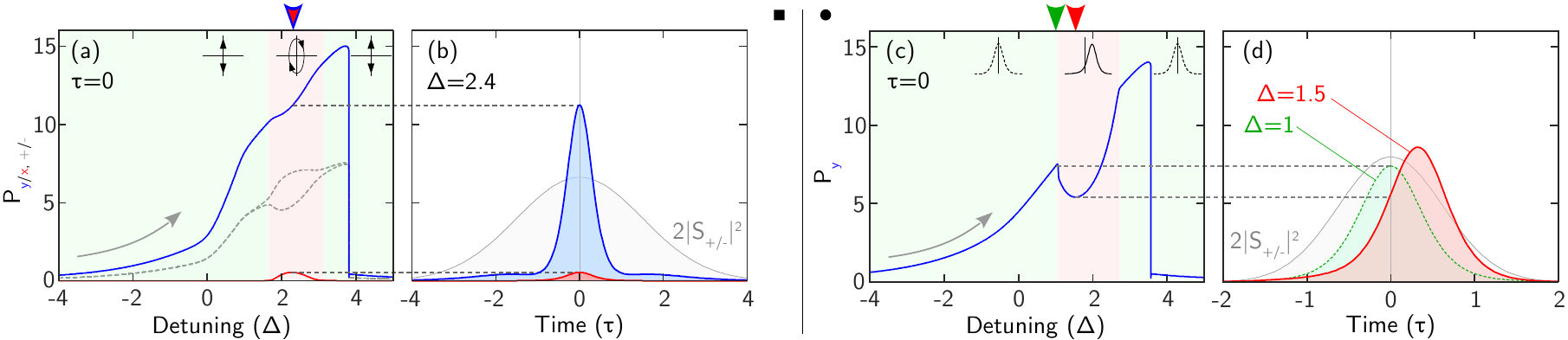}
	\caption{Numerically simulated evolution of the intracavity field when polarization (a,b) or temporal (c,d) symmetry breaking occurs while scanning the pump frequency accross a resonance. (a,c) Intracavity power polarized in the $y$ (blue curve) and $x$ (red curve) directions. (Evolution of the power of the circularly polarized components are also given in (a) as dashed gray lines). The red background denotes the detuning range for which symmetries are broken. (b,d) Input pulse profiles (gray) and intracavity pulse profiles (blue, red and green) at particular times in the scanning pointed out by coloured arrows. (a,b)  $\tau_0 = 3,\, S_0 = \sqrt{3.3}$. (c,d)  $\tau_0 = 1.25,\, S_0 = 2$. Animated version of the figure available online \cite{Viz2}.}
	\label{fig:2}
\end{figure*}

\begin{figure}[htpb]
	\centering
	\includegraphics[width=\linewidth]{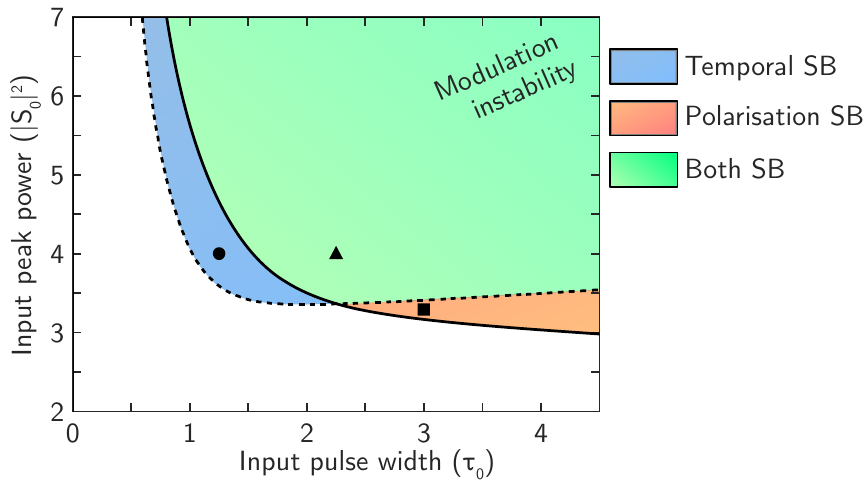}
	\caption{Chart illustrating the different domains of symmetry breaking in the parameter space of normalized input peak power $|S_0|^2$ and pulse duration $\tau_0$ with input fields of the form $S_\pm(\tau) = S_0 \exp[-(\tau/\tau_0)^2]$. The square, circle and triangle markers indicate the sets of parameters used in Fig.~\ref{fig:2}(a,b), Fig.~\ref{fig:2}(c,d), and Fig.~\ref{fig:4} respectively. SB: symmetry breaking.}
	\label{fig:3}
\end{figure}

First consider a configuration where only PSB occurs. By numerically integrating Eq.~(\ref{eq:LLEc}) we find that this is the case when scanning through the resonance with pump parameters as follows: $\tau_0 = 3,\, S_0 = \sqrt{3.3}$. Corresponding results are presented in Fig.~\ref{fig:2}(a,b) (square marker in Fig.~\ref{fig:3}). The evolution of the power of each polarization component in the orthogonal basis (coloured solid lines) and circular basis (gray dashed lines) at $\tau = 0$ (maximum of the input field) when increasing $\Delta$ is plotted in Fig.~\ref{fig:2}(a). One recognises the characteristic triangular shape with a peculiar increase of the slope around $\Delta = 0$ which marks the rising of the single peak structure inside the cavity. While $\Delta < 1.5$ the field remains linearly polarized along the $y$-axis as can be inferred from the fact that $P_x = |E_x|^2 = 0$ or equivalently by the fact that the power of the two circularly polarized components are equal. For $1.5 < \Delta < 3$, the polarization symmetry is broken and an orthogonally polarized field is generated at the expense of the $y$-component. Above $\Delta = 3$, symmetry is recovered before the system jumps out of the resonance.

\noindent In terms of circularly polarized fields, this translates into a ``bubble'' shape that is qualitatively similar to the one observed in microresonators pumped by counter-propagating fields \cite{del_bino_symmetry_2017, del_bino_microresonator_2018}. In the broken symmetry region (red background), the intracavity field exhibits an elliptical polarization and consists of a single pulse significantly shorter than the input pulse as can be seen in Fig.~\ref{fig:2}(b). The case illustrated here corresponds to a detuning scanning rate of $2\times10^{-4}$ rad/round-trip but we checked that the scenario remains qualitatively the same regardless of this value.

For shorter input pulse duration, we observed the occurrence of temporal symmetry breaking (TSB) without any sign of PSB. This is illustrated in Fig.~\ref{fig:2}(c,d) (circle marker in Fig.~\ref{fig:3}) which is the same as Fig.~\ref{fig:2}(a,b) except for the different values of the pump parameters $\tau_0 = 1.25,\, S_0 = 2$. Here, there exists a range of detuning (red background in Fig.~\ref{fig:2}(c)) for which the peak of the intracavity field is shifted with respect to the pump. This translates into a clear dip in the evolution of the power at $\tau = 0$ in Fig.~\ref{fig:2}(c). We note that a similar evolution can be obtained when the input power is swept while keeping the detuning fixed \cite{xu_experimental_2014, rossi_spontaneous_2016}. The manifestation of TSB is shown in Fig.~\ref{fig:2}(d) and we emphasize that for this particular iteration of the simulation the pulse is shifted toward positive values of $\tau$ but that owing to the spontaneous nature of the process, a shift of equal magnitude toward negative values could have occured.

In order to get further insight into the occurence of each symmetry breaking process, we performed the same numerical integrations of Eq.~(\ref{eq:LLEc}) over a large range of pump parameters ($\tau_0,\, S_0$). The results are summarized in Fig.~\ref{fig:3} which illustrates the different domains over which each process appears. A first observation is that both processes require an increasingly large input peak power to take place spontaneously when the normalized pulse duration $\tau_0$ is reduced below 1.5. We point out that in this configuration, this is typically the duration of a cavity soliton \cite{hendry_spontaneous_2018}. Secondly, we notice that the threshold for the onset of PSB (solid line) monotonically decreases as $\tau_0$ is increased. This is qualitatively similar to the results reported in Ref.~\cite{garcia-mateos_passive_1997} in the normal dispersion regime although the physics is fundamentally different: In Ref.~\cite{garcia-mateos_passive_1997}, the threshold for long pulse duration tends toward a minimum that can be determined by looking at the homogeneous stationary solutions of Eq.~(\ref{eq:LLEc}) (this would be $|S^{\mathrm{th}}_\pm|^2 = 8/\sqrt{3} \approx 4.6$ in our notation). On the other side, we found that PSB can actually occur below this threshold in the anomalous dispersion regime as a result of the buildup of MI. Indeed, the threshold can ultimately be expressed in terms of normalized intracavity power ($P^{\mathrm{th}}_\pm = |E^{\mathrm{th}}_\pm|^2 = 3$) which is locally more easily exceeded when MI kicks in. See Supplemental Material for the derivation of the thresholds and a discussion on the role of MI. Thirdly, the threshold for TSB (dashed line) appears to be minimum for a value of $\tau_0$ close to $2$ and then rises again as it crosses the PSB threshold. This latter feature can be ascribed to the fact that when PSB sets in, the peak intracavity power of the dominant polarization component ($y$) is reduced, which hinders TSB. Finally, both thresholds are exceeded over a large portion of the parameter space (in green). We should however reemphasize that the breakup of the intracavity field into multiple peaks through MI can occur over this region which does not permit an unambiguous identification of the TSB \cite{xu_experimental_2014}. We thus focus on the behaviour of the system close to thresholds. Also, a periodic evolution of the fields can be encountered above a certain threshold, corresponding to a Hopf bifurcation that we do not address in this work \cite{schmidberger_multistability_2014}. 

\begin{figure}[htbp]
	\centering
	\includegraphics[width=\linewidth]{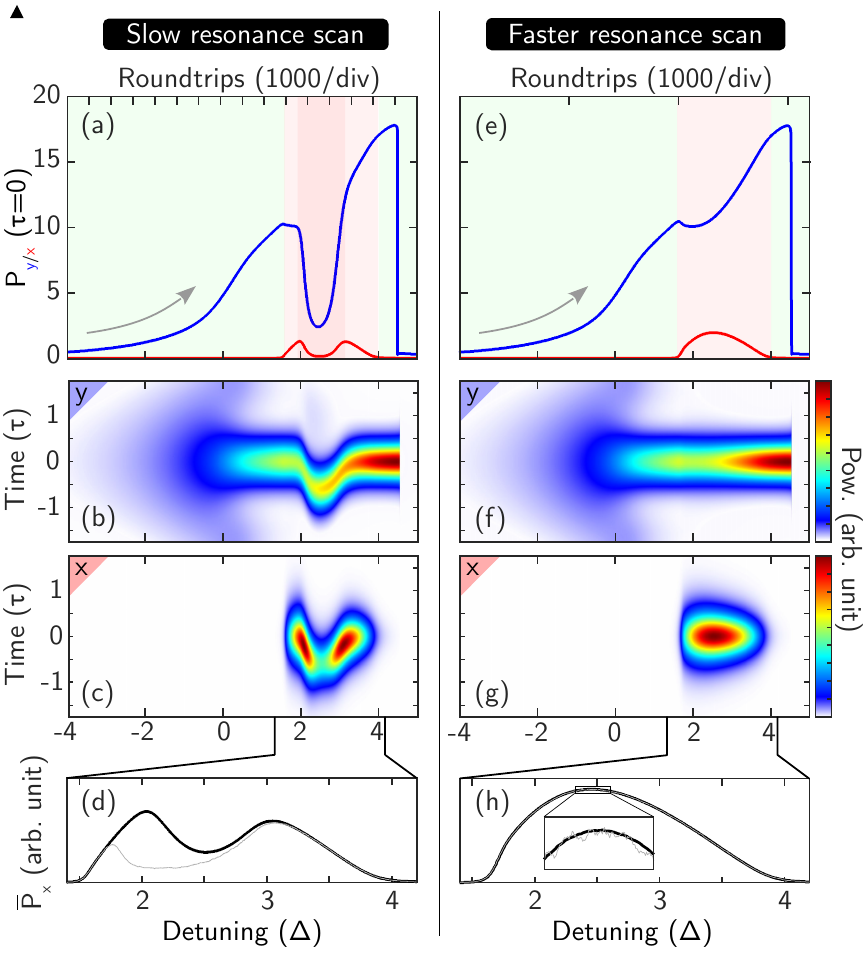}
	\caption{Dynamics of the interplay between polarization and temporal symmetry breaking for two different detuning scanning rates. (a, e) Evolution of the powers of both polarization components in the linear basis at $\tau = 0$. The red background denotes the detuning range for which one or both symmetries are broken. (b-c,f-g) 2D color plots of the evolution of the intracavity pulse profile of each polarization component. (d,h) Evolution of the average power of the $x$-component over the broken symmetry region for a certain level of power noise (black curves) and for another two orders of magnitude greater (gray). Detuning scanning rate is $5.6 \times 10^{-4}$ rad/round-trip for the left column and $2.8 \times 10^{-3}$ rad/round-trip for the right one. In both cases, $\tau_0 = 2.25,\, S_0 = 2$. Animated version of the figure available online \cite{Viz4}.}
	\label{fig:4}
\end{figure}

We now focus on the dynamics of the system for a set of parameters lying in the region where both symmetries can be broken ($\tau_0 = 2.25,\, S_0 = 2$, triangle in Fig.~\ref{fig:3}). As can be expected, an interplay between the two mechanisms takes place and controlling the dynamics of the pump field can lead to subtantially different states of the intracavity field. Indeed, we show in Fig.~\ref{fig:4} the result of two identical simulations except for the value of the scanning rate. The left column presents the dynamics of the system when the detuning is scanned at a rate of $5.6 \times 10^{-4}$ rad/round-trip. The corresponding evolution of the intracavity pulse profile of the two orthogonally polarized components is given as color plots in Fig.~\ref{fig:4}(b,c). Similarly to the previous cases, the intracavity field self-organizes into an intense pulse via MI and the system remains in a symmetric state until $\Delta \approx 1.5$ as can be inferred by both the absence of an $x$-polarized component (Fig.~\ref{fig:4}(c)) and the symmetric shape of the $y$-polarized component (with respect to $\tau$, Fig.~\ref{fig:4}(b)). The PSB occurs first, visible by the sudden increase of $P_x$ at the expense of $P_y$ for $1.5 < \Delta < 2$, rapidly followed by TSB. The latter is responsible for the rapid shift of the peak of the intracavity field toward negative fast times in this case. This in turn reduces the power of the $x$-polarized component translating into an apparent mitigation of the PSB. At this point (and for $2 < \Delta < 3$) both symmetries are simultaneously broken. Further on in the scan, the temporal asymmetry reduces and the power of the $x$-polarized component rises again until it finally vanishes before the system jumps out of resonance. The result of the same simulation performed with a scanning rate five times greater is illustrated in the right column of Fig.~\ref{fig:4}. The scenario here is in all aspects similar to the one highlighted in Fig.~\ref{fig:2}(a,b), i.e.\ showcasing PSB only over a limited range of detuning: Although TSB can potentially occur with these pump parameters, the rather fast scanning rate does not allow the process to set in. We verified that this observation is independent of the power noise level (which could eventually trigger the spontaneous TSB quicker) by performing additional simulations with increased values of the latter. Figures \ref{fig:4}(d, h) show the evolution of the average power of the $x$-polarized component over the entire broken symmetry region for two power noise levels and for each scanning rates respectively. In the case of slow scanning (Fig.~\ref{fig:4}(d)), the interplay between the two processes is significantly modified when increasing the noise level but we checked that the overall dynamics is qualitatively preserved up to levels for which the input pulse's shape is significantly degraded (typically $10\%$ of the peak power). In the case of faster scanning (Fig.~\ref{fig:4}(h)), no sign of TSB is observed regardless of the noise level, only an increasing fluctuation of the average power (see inset).

We point out that the results presented here were obtained by integrating coupled LLEs (i.e.\ in the context of the mean field model), however, we verified that numerical simulations of the full cavity map coupled equations exhibit the same features (See Supplemental Material for the equations). Also, we restrict the study to Gaussian input pulses for simplicity but our results are expected to be valid for any shape of amplitude modulation provided that only one pulse is generated by the spontaneous breakup of the input field via MI.

To conclude, we have studied a conceptually simple configuration of optical ring resonator consisting of an isotropic medium with Kerr nonlinearity synchronously pumped by short pulses. Independently, the cross-phase modulation coupling between the two circular polarization components of opposite handedness and the short pulse pumping are responsible for the occurence of polarization and temporal symmetry breaking respectively. For a certain range of pump parameters and detuning both mechanisms can coexist and a complex dynamical interplay is shown. This work makes, to our knowledge, a first link between two actively studied phenomena \cite{hendry_spontaneous_2018, hansson_modulational_2018} and might be of high relevance for future applications such as efficient pulse-pumped optical frequency comb generation, resonator-based sensor technologies, and all-optical logic gates.

\begin{acknowledgments}
	\noindent \textbf{Funding.} H2020 Marie Sklodowska-Curie Actions (MSCA) (748519, CoLiDR); National Physical Laboratory Strategic Research; H2020 European Research Council (ERC) (756966, CounterLight); Engineering and Physical Sciences Research Council (EPSRC)
\end{acknowledgments}

\pagebreak
\renewcommand{\theequation}{S\arabic{equation}}
\setcounter{equation}{0}
\onecolumngrid
	
	\begin{center}
		\begin{large}
			\textit{Supplemental material to the article:\\}
			\textbf{Interplay of Polarization and Time-Reversal Symmetry Breaking \\in Synchronously Pumped Ring Resonators}
		\end{large}

	\end{center}
	
	\section*{Introduction}
	
	This supplemental material is organized as follows: in the first section we give the main steps in the derivation of the mathematical model used in the manuscript (Eq.~(1)) starting from a general system of two coupled Ikeda-like maps. In the second section, insight into the onset of polarization symmetry breaking is obtained by a simple analysis of the model and the impact of modulation instability is discussed.
	
	\section*{Coupled Ikeda map equations and normalization}
	
	We provide in this section details regarding the derivation of the system of coupled Lugiato-Lefever equations (LLEs) in Eq.~(1) starting from two coupled full cavity (Ikeda-like) maps. The latter take the form of four coupled equations:

	\begin{equation}
		\left\{  \begin{array}{l}
			i\dfrac{\partial U_\pm}{\partial Z} + \left[\gamma \dfrac{1 - B}{2} |U_\pm|^2 + \gamma \dfrac{1 + B}{2} |U_\mp|^2 - \dfrac{\beta_2}{2} \dfrac{\partial^2 }{\partial T^2} \right] U_\pm = 0 \\[20pt]
			U_{\pm}^{(m+1)}(Z = 0, T) = \rho U_\pm^{(m)}(Z = L, T) e^{i\Phi_0} + \theta U_\pm^{\mathrm{in}} (T)
		\end{array} \right.
		\label{eq:ikeda}
	\end{equation}
	
	\noindent where $U_{\pm}$ ($U_\pm^{\mathrm{in}}$) is the left/right circularly polarized component of the intracavity (input) field envelope respectively, $Z$ and $T$ are the dimensional longitudinal coordinate along the ring and fast time defined in the reference frame traveling at the group velocity of the pump respectively, $\gamma = n_2 \omega_0/c A_{\mathrm{eff}}$ is the nonlinear coefficient with $n_2$ the nonlinear refractive index, $\omega_0$ the pump frequency and $A_{\mathrm{eff}}$ the effective area of the transverse mode. $\beta_2$ is the group velocity dispersion, $B$ is a constant related to the isotropic nonlinear medium of the cavity, and $\rho$ and $\theta$ are the amplitude reflection and transmission coefficients of the resonator. Finally, $\Phi_0$ is the linear phase accumulated over one round-trip. The first line corresponds to two nonlinear Schr\"odinger equations (NLSEs) taking into account cross-phase modulation between the two polarization modes. The second line accounts for the cavity boundary conditions applied to the two polarization modes that provide the link between the intracavity fields at the beginning of round-trip $m+1$ on one side and the intracavity fields at the end of round-trip $m$ and the input fields on the other.
	
	The linear cavity detuning is defined as $\delta_0 = 2k\pi - \Phi_0$ ($k \in \mathbb{Z}$) and the round-trip time as $t_R = nL/c$ where $L$ is the length of the ring resonator. Under the conditions of small detuning and high finesse ($\delta_0, \theta \ll 1$), (\ref{eq:ikeda}) can be reduced to the following set of two coupled LLEs \cite{lugiato_spatial_1987, haelterman_dissipative_1992}:
	
	\begin{equation}
		\frac{\partial U_\pm}{\partial Z'} = \left\{ - \frac{\theta^2}{2L} - i\frac{\delta_0}{L} + i \left[ \gamma \frac{1 - B}{2} |U_\pm|^2 + \gamma \frac{1 + B}{2} |U_\mp|^2 - \frac{\beta_2}{2} \frac{\partial^2 }{\partial T'^2} \right] \right\}  U_\pm + \frac{\theta}{L} U^{\mathrm{in}}_{\pm}(T')
		\label{eq:LLEdim}
	\end{equation}
	
	\noindent where new fast time and longitudinal coordinates are introduced as $T' = m t_R$ and $t_R \dfrac{\partial }{\partial T'} = L \dfrac{\partial }{\partial Z'}$. In the main manuscript we refer to a normalized version of Eqs.~(\ref{eq:LLEdim}) which is obtained by the following change of variables:
	
	\begin{equation}
		\left\{  \begin{array}{l}
			\alpha = \dfrac{\theta^2}{2L}\\[10pt]
			\Delta = \dfrac{\delta_0}{\alpha L}\\[10pt]
			\eta = \dfrac{\beta_2}{|\beta_2|}\\[10pt]
			\tau = \sqrt{\dfrac{\alpha}{|\beta_2|}} T'\\[10pt]
			z = \alpha Z'\\[10pt]
			E_\pm = \sqrt{\dfrac{\gamma}{\alpha}} U_\pm\\[10pt]
			S_\pm = \sqrt{\dfrac{\gamma \theta^2}{\alpha^3 L^2}} U^{\mathrm{in}}_\pm
		\end{array} \right.
		\label{eq:var_change}
	\end{equation}
	
	\noindent where $\alpha$ represents the overall losses, $\eta$ the sign of the group velocity dispersion, $\tau$ and $z$ the normalized fast time and longitudinal coordinates respectively, $E_\pm$ and $S_\pm$ the normalized intracavity and input fields respectively. In this notation, $\Delta$ is the normalized detuning in units of half the linewidth at half maximum of the linear resonance. Finally, the normalized system of two LLEs is:
	
	\begin{equation}
		\frac{\partial E_\pm}{\partial z} - i \left[\frac{1 - B}{2} |E_\pm|^2 + \frac{1 + B}{2} |E_\mp|^2 - \frac{\eta}{2} \frac{\partial^2 }{\partial \tau^2} \right] E_\pm + (1 + i \Delta) E_\pm = S_\pm (\tau)
		\label{eq:LLEadim}
	\end{equation}
	
	\vspace{2em}
	
	\section*{Derivation of the polarization symmetry breaking thresholds}
	
	Polarization symmetry breaking (PSB) thresholds in terms of detuning ($\Delta^{\mathrm{th}}$), intracavity power ($|E^{\mathrm{th}}_\pm|^2$), and input power ($|S^{\mathrm{th}}_\pm|^2$) can be determined by a simple analysis of Eqs.~(\ref{eq:LLEadim}) (Eqs. (1) in the main manuscript) in the homogeneous case (i.e. not considering the breaking of the intracavity field into pulses as a result of the modulation instability (MI) process). The aim of this supplementary section is to provide the derivation of these threshold values (The method followed here is similar to the one described in Refs. \cite{kaplan_directionally_1982, haelterman_polarization_1994}). We then point out that, in the presence of MI, symmetry breaking can occur for parameters below these thresholds.\\
	
	Conditions for the occurence of PSB in the homogeneous case are determined by considering the stationary $\left(\dfrac{\partial E_\pm}{\partial z} = 0 \right)$ and homogeneous $\left( \dfrac{\partial^2 E_\pm}{\partial \tau^2} = 0 \right)$ solutions of (\ref{eq:LLEadim}). These are given by:
	
	\begin{equation}
		|E_\pm|^2 \left[ 1 + \left( \dfrac{1}{3} |E_\pm|^2 + \dfrac{2}{3} |E_\mp|^2 - \Delta\right)^2 \right] = |S_\pm|^2
		\label{eq:Lorentz}
	\end{equation}
	
	\noindent which can be seen as a set of two coupled nonlinear Lorentzian resonances ($|E_\pm|^2 = f(\Delta)$) for the two polarization modes. Here we fixed $B = 1/3$ as it is the case in the main manuscript. PSB sets in under symmetric pumping (i.e. $|S_+|^2 = |S_-|^2$). Under this condition, the difference of the two Equations in (\ref{eq:Lorentz}) is a cubic equation that can be expressed as follows:
	
	\begin{equation}
		\underbrace{\left( |E_-|^2 - |E_+|^2 \right)}_{\text{(Sym)}} \underbrace{\left[\frac{|E_+|^4 + |E_+|^2 |E_-|^2 + |E_-|^4}{9} - 2\Delta \frac{|E_+|^2 + |E_-|^2}{3} + 1 + \Delta^2 \right]}_\text{(SymBr)} = 0
		\label{eq:cubic}
	\end{equation}
	
	The first factor labelled (Sym) denotes symmetric intracavity states (i.e. $|E_+|^2 = |E_-|^2$) whereas the second factor labelled (SymBr) characterizes states with broken symmetry (i.e. $|E_+|^2 \neq |E_-|^2$). When the system is on the cusp of symmetry breaking, we can write $|E_+|^2 = |E_-|^2 = |E^{\mathrm{SB}}_\pm|^2$ and (SymBr) becomes:
	
	\begin{equation}
		|E^{\mathrm{SB}}_\pm|^4 - 4 \Delta |E^{\mathrm{SB}}_\pm|^2 + 3(1 + \Delta^2)= 0
		\label{eq:polynom}
	\end{equation}
	
	\noindent The roots of Eq.~(\ref{eq:polynom}) are:
	
	\begin{equation}
		|E^{\mathrm{SB}}_{\pm\pm}|^2 = 2\Delta \pm \sqrt{\Delta^2 - 3}
		\label{eq:roots}
	\end{equation}
	
	\noindent from which one can see that PSB can only occur above the detuning threshold $\Delta^{\mathrm{th}} = \sqrt{3}$ which also happens to be the threshold of scalar bistability \cite{haelterman_polarization_1994} (The first $\pm$ on the left-hand side of (\ref{eq:roots}) relates to the two circularly polarized components whereas the second one relates to the two roots). The minimum threshold for PSB in terms of intracavity power is easily found from Eq.~(\ref{eq:roots}) to be $|E^{\mathrm{th}}_\pm(\Delta = 2)|^2 = 3$. Similarly, substituting Eq.~(\ref{eq:roots}) to (\ref{eq:Lorentz}) with $|E_+|^2 = |E_-|^2 = |E^{\mathrm{SB}}_\pm|^2$, one finds that the minimum threshold for PSB in terms of input power is $|S^{\mathrm{th}}_\pm(\Delta = 5/\sqrt{3})|^2 = 8/\sqrt{3} \approx 4.6$.
	
	A noteworthy conclusion here is that the PSB intracavity power threshold ($|E^{\mathrm{th}}_\pm|^2 = 3$) is greater than the MI threshold $|E^{\mathrm{MI, th}}_\pm|^2 = 1$ \cite{haelterman_dissipative_1992}. This implies that in a realistic configuration (that considers non-homogeneous solutions) MI might kick in and break up a cw field into a train of pulses whose local power exceeds the PSB threshold for a pump power below the one predicted in the homogeneous case. This explains why we observe the occurence of PSB in our numerical simulations for normalized pump peak power values as low as $|S_0|^2 = 3$ (See Fig. 3 of the main manuscript toward long pulse duration $\tau_0$) which is below the ``homogeneous threshold'' $|S^{\mathrm{th}}_\pm|^2 \approx 4.6$. Note that in our simulations, we find that the PSB threshold increases for shorter pulse duration and exceeds the ``homogeneous threshold'' for $\tau_0 \lesssim 1.2$ which is consistent with the fact that this value of $\tau_0$ is typically the MI period for this pump power.

\end{document}